\documentclass[aps,twocolumn,showpacs,preprintnumbers,superscriptaddress,a4paper]{revtex4}

\usepackage{times}
\usepackage{subfig}
\usepackage{bm}
\usepackage{graphicx}
\usepackage{amsbsy}
\usepackage{amsmath}
\usepackage{amsfonts}
\usepackage{amsthm}
\usepackage{float}
\usepackage{color}
\usepackage{amssymb}

\begin{document}
\theoremstyle{plain}
\newtheorem{theorem}{Theorem}
\newtheorem{lemma}[theorem]{Lemma}
\newtheorem{corollary}[theorem]{Corollary}
\newtheorem{proposition}[theorem]{Proposition}
\newtheorem{conjecture}[theorem]{Conjecture}

\theoremstyle{definition}
\newtheorem{definition}[theorem]{Definition}

\title{Quantification of Entanglement of Teleportation 
in Arbitrary Dimensions}
\author {Sk Sazim}
\affiliation{Institute of Physics, Sainik School Post,
Bhubaneswar-751005, Orissa, India}
\author{Satyabrata Adhikari}
\affiliation{Indian Institute of Technology Rajasthan, Jodhpur-342011, India}
\author{Subhashish Banerjee}
\affiliation{Indian Institute of Technology Rajasthan, Jodhpur-342011, India}
\author{Tanumoy Pramanik}
\affiliation{S. N. Bose National Centre for Basic Sciences, Salt Lake, Kolkata-700098, India}

\begin{abstract}
We study bipartite entangled states in arbitrary dimensions and obtain different bounds for the entanglement measures in
terms of teleportation fidelity. We find that there is a simple relation between
negativity and teleportation fidelity for pure states but for mixed states, an upper bound is obtained for negativity in terms
of teleportation fidelity using convex-roof extension negativity (CREN). However, with this it is not clear how to distinguish
betweeen states useful for teleportation and positive partial transpose (PPT) entangled states. Further, there exists a strong conjecture in the literature  that all 
PPT entangled states, in $3 \times 3$ systems, have Schmidt rank two.
This motivates us to develop measures capable of identifying states useful for teleportation and dependent
on the Schmidt number. We thus establish various relations between teleportation fidelity and entanglement
measures depending upon Schmidt rank of the states. These relations and bounds help us to determine the
amount of entanglement required for teleportation, which we call the ``Entanglement of Teleportation''. These bounds are used to determine the teleportation fidelity 
as well as the entanglement required for teleportation of  states modeled by a two qutrit mixed system,  as well as two qubit open quantum systems.
\end{abstract}

\maketitle
\section{Introduction}
Entanglement \cite{EPR} lies at the heart of quantum mechanics. For a long time it was considered synonymous with
quantum correlations but  is now regarded as a subset of quantum correlations, see for example \cite{vedral}. Entanglement plays a pivotal role in various information processing tasks, including, among others,  
quantum teleportation \cite{CHB}, super dense coding \cite{CHB2}, remote state
preparation \cite{AKP}, secret sharing \cite{MH}, and quantum cryptography \cite{NG}.

In quantum teleportation, using entangled states as resource, it is possible to transfer quantum information from an unknown
qubit to another one placed at a distance. Thus, one of the party, say, Alice makes a two
qubit measurement on her qubit and the unknown state in Bell basis, and sends the measurement results through a
classical channel to the second party, say, Bob (who is located away from Alice). Accordingly,
Bob makes appropriate unitary transformations to obtain the desired state. Thus the ability of
teleporting an unknown state depends on the nature of entanglement of the resource state and 
is called teleportation fidelity.

The situation is very straight forward when we have an unknown qubit to send with the help of a pure entangled state
as a resource. However, it is more involved when we have mixed entangled states as a medium of
teleportation. For a general two qubit density matrix
$\rho=\frac{1}{4}[I \otimes I+\sum_{i}r_i(\sigma_i \otimes I)+\sum_{j} s_j(I \otimes \sigma_j)+
\sum_{i,j} t_{ij}(\sigma_i \otimes\sigma_j)]$, the teleportation fidelity is a function of the eigenvalues of correlation
matrix $T=[t_{ij}]$. Similarly, when we go from  qubits to higher dimensional bipartite states the
teleportation fidelity is expressed in terms of the singlet fraction of the state. The relation between optimal
teleportation fidelity $F(\rho)$ and maximal singlet fraction$f(\rho)$ in a $d\otimes d$ system, if one performs
quantum teleportation with the state $\rho$, is \cite{horodecki1}
\begin{eqnarray}
F(\rho)= \frac{df(\rho)+1}{d+1} \label{T18}.
\end{eqnarray}
Here the singlet fraction is defined as, $f(\rho)=\max_{|\psi\rangle} \langle\psi|\rho|\psi\rangle$, and
$|\psi \rangle$ is a maximally entangled state in $d \otimes d$.
If $f(\rho)> \frac{1}{d}$ then the parties can perform quantum teleportation with the average fidelity of the
teleported qubit exceeding the classical limit $\frac{2}{d+1}$.

In bipartite two qubit states it is known that the total amount of entanglement present in the resource state is
useful for teleportation. Then one can ask the following question: How much entanglement is necessary for teleporting an
unknown state when we have a bipartite state in arbitrary dimensions? To answer this question one has to quantify
the entanglement and find out for what range of entanglement the state can be used as a resource for teleportation. In other words, one needs to establish a relationship between the amount of entanglement and teleportation fidelity. 
We establish various relations between teleportation fidelity and entanglement measures depending upon Schmidt rank \cite{ES,rank} of the states. These relations and bounds help us to answer the above question. Given an arbitrary two-qudit state 
with Schmidt rank upto three we can predict its utility as a resource for teleportation.  

Negativity \cite{peres96} is a measure of entanglement of a bipartite quantum state described by the density operator $\rho$ 
and is formally defined in $d \times d$ systems as
\begin{eqnarray}
N(\rho)=\frac{||\rho^{T_{A}}|| -1}{d-1},
\end{eqnarray}
where $\rho^{T_{A}}$ is a partial transpose of $\rho$ with respect to the system $A$ and $||.||$ denotes the trace norm.
Negativity fails to distinguish separable states from PPT
entangled states, that is, bound entangled states. This difficulty can be overcome by the use of convex-roof extension of negativity (CREN) \cite{lee03}.

There is strong evidence in $3 \times 3$
systems that bound entangled states exist only for states with Schmidt rank two \cite{sanpera01}.  Thus the entangled states with Schmidt rank three would,
presumably, be useful for teleportation. This provides a strong motivation to study the states with Schmidt rank three from the perspective of
teleportation fidelity.  The Schmidt number is a very useful entanglement measure \cite{sperling11}. 
In the literature, there exists different kinds of entanglement measures, expressed in terms of Schmidt numbers, suitable for quantification of the 
amount of entanglement present in the system. 

 We  quantify the amount of entanglement present in the resource state to find out
the bounds within which these states can be useful for teleportation.  Thus, we obtain relations
connecting  entanglement measures with teleportation fidelity using CREN as well as  singlet fraction, expressed in terms of
Schmidt coefficients. Our results are
obtained for arbitrary dimensional bipartite states with at most three non vanishing Schmidt coefficients. We implement our
results to detect mixed states useful for teleportation. A monotonous relation between entanglement and teleportation fidelity
in mixed two qudit systems could be expected from \cite{ben}, where a monotonous connection between entanglement and singlet fraction, and hence
teleportation fidelity, was established for two qubit mixed states.

The plan of the paper is as follows. 
In Section 2, we study the relation between negativity and teleportation fidelity for pure as well as mixed systems.
Based on our conclusions from Section 2, we establish a relation between singlet fraction and
different types of entanglement measures for arbitrary dimensional pure two qudit system with a maximum of three Schmidt
coefficients in Section 3.  Then we study the bounds of teleportation fidelity
and entanglement measures for two special cases, i) arbitrary dimensional pure bipartite state with two Schmidt coefficients,
and ii) arbitrary dimensional pure bipartite state with three Schmidt coefficients. These results are used in section 3 (B) to 
arbitrary dimensional mixed bipartite systems with Schmidt coefficients two and three. In section 4, we apply our results on
examples of mixed states, in particular, two qutrit mixed state with Schmidt rank two,  and two qubit mixed states generated dynamically by an open system model.  Finally, we conclude in section 5.

\section{Relation between Negativity and Teleportation fidelity for $d\otimes d$ systems}
Here we study the relation between negativity and teleportation fidelity for pure as well as mixed systems.

\subsection{Pure Systems}
Let $H_{A}$ and $H_{B}$ be two Hilbert spaces each with dimension
$d$. In $d\otimes d$ system, any pure state $|\Psi\rangle$ can be
expressed as
\begin{eqnarray}
|\Psi\rangle=\sum_{i=1}^{d}\sqrt{\lambda_{j}}|j\rangle|j\rangle.
\label{purestate}
\end{eqnarray}
The negativity of the state $|\Psi\rangle$ is defined as
\begin{eqnarray}
N(|\Psi\rangle)=\frac{2}{d-1}\sum_{i<j}\sqrt{\lambda_{i}\lambda_{j}}.
\label{negativity}
\end{eqnarray}
The singlet fraction for any pure state in $d\otimes d$ system is
given by
\begin{eqnarray}
f(|\Psi\rangle)=\frac{1}{d}\left(\sum_{i=1}^{d}\sqrt{\lambda_{i}}\right)^{2}.
\label{singletfraction}
\end{eqnarray}
The relation between negativity and singlet fraction is given by \cite{horo99}
\begin{eqnarray}
N(|\Psi\rangle)=\frac{df(|\Psi\rangle)-1}{d-1}.\label{relbetsingletfractionneg}
\end{eqnarray}
In terms of teleportation fidelity,
Eq. (\ref{relbetsingletfractionneg}) reduces to
\begin{eqnarray}
F(|\Psi\rangle)=\frac{2}{d+1}+\frac{(d-1)N(|\Psi\rangle)}{d+1}.
\label{relbettelfidneg}
\end{eqnarray}
Therefore, it follows that every entangled pure state in a $d\otimes
d$ system is useful for teleportation.

\subsection{Mixed Systems}
A bipartite mixed state described can be described by the density operator $\rho$
\begin{eqnarray}
\rho=\sum_{i} p_{i}|\Psi_{i}\rangle\langle
\Psi_{i}|\label{mixedstate}.
\end{eqnarray}
The negativity of the mixed state $\rho$ can be extended from the
pure state by means of convex roof, that is, convex-roof extended negativity (CREN) \cite{lee03}:
\begin{eqnarray}
N(\rho)=\min_{\{p_{i},|\Psi_{i}\rangle\}}\sum_{i}p_{i}N(|\Psi_{i}\rangle). \label{mixedstate}
\end{eqnarray}
The upper bound of the negativity of the mixed state $\rho$ can be
expressed in terms of the singlet fraction as
\begin{eqnarray}
N(\rho)&&=\min_{\{p_{i},|\Psi_{i}\rangle\}}\sum_{i}p_{i}N(|\Psi_{i}\rangle)\nonumber{}\\&&\leq
\sum_{i}p_{i}N(|\Psi_{i}\rangle)\nonumber{}\\&&=\frac{d}{d-1}\sum_{i}p_{i}f(|\Psi_{i}\rangle)-\frac{1}{d-1}. \label{bound1}
\end{eqnarray}
In terms of teleportation fidelity, the bound on negativity is
\begin{eqnarray}
N(\rho)\leq \frac{d+1}{d-1}\sum_{i}p_{i}F(|\Psi_{i}\rangle)-\frac{2}{d-1}. \label{bound2}
\end{eqnarray}
The above inequality (\ref{bound2}) measures the upper bound of
entanglement contained in the mixed state $\rho$. From this, it is clear that
CREN detects both PPT bound entangled states as well as states useful
for teleportation. However, it is not clear how to distinguish between these two classes of states.
Further, there exists a strong conjecture in the literature \cite{sanpera01} that all PPT
entangled states, in $3 \times 3$
systems, have Schmidt rank two.
This motivates us to develop measures capable of identifying states useful for teleportation and dependent
on the Schmidt number.

\section{Relation between singlet fraction and different entanglement measures for two qudit system with
three Schmidt coefficients}
In this section we obtain an explicit relation that will connect entanglement monotones with singlet
fraction for a two qudit system of arbitrary dimension. We obtain results
in $d\otimes d$ systems with two and three
non zero Schmidt coefficients. 

\subsection{Bounds on entanglement measures for pure two qudit systems useful for teleportation}
Let us consider a bipartite $d \otimes d$ system in which three Schmidt coefficients are non zero. Without any
loss of generality we assume that the first three Schmidt coefficients are non zero.
Any pure two qudit system with three non zero Schmidt coefficients $\lambda_1$, $\lambda_2$ and $\lambda_3$ can
be written in Schmidt decomposition form as,
\begin{eqnarray}
|\Psi^d\rangle =\sqrt{\lambda_1} |00\rangle + \sqrt{\lambda_2} |11\rangle + \sqrt{\lambda_3} |22\rangle,
\end{eqnarray}
with the Schmidt coefficients summing to one, i.e.,   $\lambda_1+\lambda_2+\lambda_3=1$.
To quantify the amount of entanglement in $|\Psi^d\rangle$ we consider two different entanglement measures
$E^{(d,2)} (|\Psi^{d}\rangle)$ and $E^{(d,3)} (|\Psi^d\rangle)$ which can be defined as \cite{Gour},
\begin{eqnarray}
E^{(d,2)}(|\Psi^{d}\rangle)= \sqrt{\frac{2 d}{d-1} (\lambda_1 \lambda_2+\lambda_2 \lambda_3+\lambda_1 \lambda_3)},
\label{EM1}
\end{eqnarray}
\begin{eqnarray}
E^{(d,3)}(|\Psi^{d}\rangle)= \Big(\frac{6 d^2}{(d-1)(d-2)}\Big)^{\frac{1}{3}} (\lambda_1 \lambda_2 \lambda_3)^{\frac{1}{3}}.
\label{EM2}
\end{eqnarray}

\noindent Here $E^{(d,2)}(|\Psi^{d}\rangle)$  and $E^{(d,3)}(|\Psi^{d}\rangle)$ denote entanglement measure for a $d\otimes d$ dimensional pure system
defined by taking the sum of the product of the Schmidt coefficients taken two or three at a time, respectively.  We note that for a Schmidt rank
two state, $E^{(d,3)} (|\Psi^{d}\rangle)=0$ but $E^{(d,2)} (|\Psi^{d}\rangle)\neq 0$.

The singlet fraction for the state $|\Psi^{d}\rangle$ is defined as
\begin{eqnarray}
f(|\Psi^{d}\rangle)=\max_{|\Phi\rangle} |\langle\Phi |\Psi^{d}\rangle|^2,
\label{fra}
\end{eqnarray}
where the maximum is taken over all maximally entangled states $|\Phi\rangle$ in $d\otimes d$ systems.
The singlet fraction $f(|\Psi^{d}\rangle)$ for pure state $|\Psi^{d}\rangle$ can also be expressed in terms
of Schmidt coefficients \cite{Horo} as
\begin{eqnarray}
f(|\Psi^{d}\rangle)=\frac{1}{d} \Big(\sqrt{\lambda_1}+\sqrt{\lambda_2}+\sqrt{\lambda_3}\Big)^2.
\label{SF}
\end{eqnarray}
Expanding the the right hand side part of Eq. (\ref{SF}) and using $\lambda_1+\lambda_2+\lambda_3=1$, we get
\begin{eqnarray}
\sqrt{\lambda_1 \lambda_2}+\sqrt{\lambda_2 \lambda_3}+\sqrt{\lambda_1 \lambda_3}= \frac{d f(|\Psi^{d}\rangle) - 1}{2}.
\label{SF1}
\end{eqnarray}
Also, we have the following  identity
\begin{eqnarray}
\lambda_1 \lambda_2+\lambda_2 \lambda_3+\lambda_1 \lambda_3=(\sqrt{\lambda_1 \lambda_2}+\sqrt{\lambda_2 \lambda_3}
+\sqrt{\lambda_1 \lambda_3})^2\nonumber\\ -2 \sqrt{\lambda_1 \lambda_2 \lambda_3} (\sqrt{\lambda_1}
+\sqrt{\lambda_2}+\sqrt{\lambda_3}).
\label{IDN1}
\end{eqnarray}
Using (\ref{EM1}), (\ref{EM2}), (\ref{SF}), (\ref{SF1}) and (\ref{IDN1}) we have
\begin{eqnarray}
(E^{(d,2)} (|\Psi^{d}\rangle))^{2} &=& \frac{d^3}{2 (d -1)} \left(f(|\Psi^{d}\rangle)-\frac{1}{d}\right)^2\nonumber\\
&-& \frac{4}{d-1} \sqrt{\frac{d(d-1)(d-2)}{6}} (E^{(d,3)} (|\Psi^{d}\rangle))^{\frac{3}{2}}\nonumber\\ &\times& \sqrt{f(|\Psi^{d}\rangle)}.
\label{E32}
\end{eqnarray}

This establishes the required relationship between the entanglement measures
$E^{(d,2)}(|\Psi^{d}\rangle)$ and $E^{(d,3)}(|\psi^{d})\rangle$ with the singlet fraction $f(|\Psi^{d}\rangle)$
for a pure two qudit system $|\Psi^{d}\rangle$ with three non vanishing Schmidt coefficients.

Next, we will consider separately the cases of states of Schmidt ranks two and three, respectively.
For purpose of clarity, in the discussions to follow, we modify the notation of the entanglement 
measures discussed above, as $E_j^{d,i}$, where $d$ stands for the $d \otimes d$ dimensional system,
$j$ indicates the Schmidt rank of the state under consideration and $i$ is the number of coefficients
taken at a time.  

\subsubsection{States with Schmidt Rank Two }

When one of the Schmidt coefficients (say, $\lambda_3$) is zero, i.e.,
$E_2^{(d,3)}(|\Psi^{d}\rangle)=0$, from Eq. (\ref{E32}), we have
\begin{eqnarray}
E^{(d,2)}_2 (|\Psi^{d}\rangle)=\sqrt{\frac{d^3}{2 (d -1)}} \left(f_2 (|\Psi^{d}\rangle)-\frac{1}{d}\right), \phantom{xxxx}
\label{E320}
\end{eqnarray}
where, $f_2(|\Psi^{d}\rangle)$ denotes the singlet fraction of Schmidt rank two state, and
$f_2(|\Psi^{d}\rangle) > \frac{1}{d}$.
If $F_2(|\Psi^{d}\rangle)$ denotes the teleportation fidelity of Schmidt rank two states, then
$E^{(d,2)}_2 (|\Psi^{d}\rangle)$ can be expressed in terms of $F_2(|\Psi^{d}\rangle)$ as
\begin{eqnarray}
E^{(d,2)}_2 (|\Psi^{d}\rangle)=\sqrt{\frac{d^3}{2 (d -1)}} \Bigg[\frac{(d+1)F_2 (|\Psi^{d}\rangle)-2}{d}\Bigg].
\end{eqnarray}
This establishes the relation between the entanglement monotone and teleportation fidelity of Schmidt rank
two states. If the state $|\Psi^{d}\rangle$ has Schmidt number two and useful for teleportation, then we
have \cite{Terhal}
\begin{eqnarray}
\frac{1}{d} < f_2(|\Psi^{d}\rangle) \leq \frac{2}{d}.
\label{f22}
\end{eqnarray}
Eq. (\ref{f22}) can be recast in terms of teleportation fidelity as
\begin{eqnarray}
\frac{2}{d+1} < F_2(|\Psi^{d}\rangle) \leq \frac{3}{ d+1}.
\label{F22}
\end{eqnarray}
Using Eq. (\ref{F22}), $E^{(d,2)}_2 (|\Psi^{d}\rangle)$ can be seen to be bounded as
\begin{eqnarray}
0< E^{(d,2)}_2 (|\Psi^{d}\rangle) \leq \sqrt{\frac{d}{2 (d -1)}}. \label{entschmidt2}
\end{eqnarray}
When the amount of entanglement lies in
the above range we can use the state for teleportation. This quantifies the entanglement required for teleportation for
a pure qudit state with two non-vanishing Schmidt coefficients.

\subsubsection{States with Schmidt Rank Three }

Next we take up sates where none of the three Schmidt coefficients are zero, i.e., $E^{(d,3)}_3 (|\Psi^{d}\rangle)\neq 0$.

Using the well known result of arithmetic mean (AM) being greater than or equal to geometric mean  (GM) on
three real quantities $\sqrt{\lambda_1 \lambda_2}, \sqrt{\lambda_1 \lambda_3}$
and $\sqrt{\lambda_2 \lambda_3}$, we have
\begin{eqnarray}
\frac{\sqrt{\lambda_1 \lambda_2}+ \sqrt{\lambda_1 \lambda_3} + \sqrt{\lambda_1 \lambda_3}}{3} \geq
\Big(\sqrt{\lambda_1 \lambda_2} \sqrt{\lambda_1 \lambda_3} \sqrt{\lambda_2 \lambda_3}\Big)^{\frac{1}{3}}. \nonumber \\
\end{eqnarray}
Using Eqs. (\ref{EM2}), and (\ref{SF1}), we have
\begin{eqnarray}
f_3(|\Psi^{d}\rangle) \geq  \frac{6}{d} \Bigg[\Big(\frac{(d-1) (d-2)}{6 d^2}\Big)^{\frac{1}{3}} E^{(d,3)}_3(|\Psi^{d}\rangle) \Bigg] + \frac{1}{d}.
\end{eqnarray}
Since, the singlet fraction $f_3(|\Psi^{d}\rangle)$ attains its maximum value unity at
$\lambda_1=\lambda_2=\lambda_3=\frac{1}{d}$, we have
\begin{eqnarray}
\frac{6}{d} \Bigg[\Big(\frac{(d-1) (d-2)}{6 d^2}\Big)^{\frac{1}{3}} E^{(d,3)}_3(|\Psi^{d}\rangle) \Bigg] + \frac{1}{d}  \nonumber \\
\leq f_3(|\Psi^{d}\rangle) \leq 1.
\label{cond3}
\end{eqnarray}
In terms of teleportation fidelity $F_3(|\Psi^{d}\rangle)$, the above inequality can be expressed as
\begin{eqnarray}
\frac{2}{d+1}+\frac{6}{d+1} \Big(\frac{(d-1) (d-2)}{6 d^2}\Big)^{\frac{1}{3}} E^{(d,3)}_3(|\Psi^{d}\rangle) \nonumber \\
\leq F_3(|\Psi^{d}\rangle) \leq 1.
\label{cond4}
\end{eqnarray}
Hence, pure entangled states with $E^{(d,3)}_3(|\Psi^{d}\rangle)$ satisfying Eq. (\ref{cond4}) and teleportation fidelity $F_3(|\Psi^{d}\rangle) >
\frac{2}{d+1}$ are Schmidt rank three states useful for teleportation.

\subsection{Bounds on entanglement measures for mixed two qudit systems useful for teleportation}

In this section we would like to answer the following questions :
(i) What is the minimum amount of entanglement needed to perform teleportation when the 
mixed state eith Schmidt rank two is used as a resource in a $d\otimes d$ system?
(ii)  What is the minimum amount of entanglement needed to perform teleportation when the 
mixed state with Schmidt rank three is used as resource in a $d\otimes d$ system?

Let us consider a mixed qudit state described by the density operator $\rho =\displaystyle\sum_{i=1}^n p_i \rho_i$,
where $\sum_{i=1}^n p_i=1$ and $\rho_i$ ($=|\psi_i\rangle\langle \psi_i|$) are composite pure states.
The singlet fraction $f(\rho)$ of the state $\rho$ is defined as
\begin{eqnarray}
f(\rho)=\max_U \langle\psi^+| U^{\dagger} \otimes \mathcal{I} \rho U \otimes \mathcal{I} |\psi^+\rangle,
\label{sf2}
\end{eqnarray}
 where  $U$ is the unitary matrix, $\mathcal{I}$ is the identity matrix and $|\psi^+\rangle=
\frac{1}{\sqrt{d}}\displaystyle\sum_{k=0}^{d-1} |kk\rangle$ represents a pure maximally entangled state.

The entanglement measure $E^{(d,2)}(|\Psi^{d}\rangle)$ and $E^{(d,3)}(|\Psi^{d}\rangle)$ given in
Eqs. (\ref{EM1}) and (\ref{EM2}) for pure states can also be defined for a mixed state $\rho$ as
\begin{eqnarray}
E^{(d,2)}(\rho)=\min \displaystyle\sum_{i=1}^n p_i E^{(d,2)}(\rho_i),
\label{emr1}
\end{eqnarray}
and
\begin{eqnarray}
E^{(d,3)}(\rho)=\min \displaystyle\sum_{i=1}^n p_i E^{(d,3)}(\rho_i).
\label{emr2}
\end{eqnarray}
Here the minimum is taken over all pure state decompositions of $\rho$.
Now one may ask a question that, like entanglement measures, does the singlet fraction $f(\rho)$
also have the property \cite{prop1}
\begin{eqnarray}
f(\rho)=\min \displaystyle\sum p_i f(\rho_i),
\end{eqnarray}
where the minimum is taken over all decomposition of $\rho$. Unfortunately, the answer is no.

\subsubsection{Two qudit mixed state of Schmidt rank two}

From Eq. (\ref{E32}), $E^{(d,2)}_2(\rho_i)$ for any bipartite pure qudit
state with Schmidt rank two $\rho_i$ whose $f_2(\rho_i)=\frac{1}{d}$, i.e., for states not useful for teleportation, we have 
\begin{equation}
  E^{(d,2)}_2(\rho_i)=0.
\end{equation}
In general for any bipartite pure qudit
state with Schmidt rank two $\rho_i$ useful for teleportation, the entanglement $E^{(d,2)}_2$ is
\begin{eqnarray}
E^{(d,2)}_2 (\rho_i)=\sqrt{\frac{d^3}{2 (d-1)}}\Big(f_2(\rho_i)-\frac{1}{d}\Big).
\label{E321}
\end{eqnarray}
Using Eqs. (\ref{emr1}) and (\ref{E321}), we have
\begin{eqnarray}
E^{(d,2)}_2(\rho)&=&\min\displaystyle\sum_{i} p_i \sqrt{\frac{d^3}{2 (d-1)}}\left(f_2(\rho_i)-\frac{1}{d}\right) \nonumber\\
  &\leq &\displaystyle\sum_{i} p_i \sqrt{\frac{d^3}{2 (d-1)}}\left(f_2(\rho_i)-\frac{1}{d}\right) \nonumber \\
  & < &  \sqrt{\frac{d}{2(d-1)}},\label{new45}
\end{eqnarray}
where the last inequality follows from an application of Eq. (\ref{f22}).
Hence, if the mixed state $\rho$ with Schmidt rank two in a $d \otimes d$ system is useful for teleportation then
\begin{eqnarray}
0 < E^{(d,2)}_2(\rho) < \sqrt{\frac{d}{2(d-1)}}.
\label{BE32}
\end{eqnarray}

\subsubsection{Two qudit mixed state of Schmidt rank three}

Using, once again, the result of arithmetic mean (AM) being greater than or equal to geometric mean  (GM) on
three real quantities $\lambda_1 \lambda_2$, $\lambda_1 \lambda_3$
and $\lambda_2 \lambda_3$ and Eqs. (\ref{EM1}), (\ref{EM2}) we obtain the following bound on $E^{(d,3)}_3(\rho)$ for two qudit mixed states with Schmidt rank three:
\begin{equation}
0 < E^{(d,3)}_3(\rho) < \Big[\frac{d(d-1)}{6}\Big]^{\frac{1}{6}} \frac{1}{(d-2)^{1/3}}. \label{schmidt3}
\end{equation}
Comparing Eqs. (\ref{schmidt3}) and (\ref{BE32}), we can see that if the entanglement lies in the range $\sqrt{\frac{d}{2(d-1)}}$ to
$\Big[\frac{d(d-1)}{6}\Big]^{\frac{1}{6}} \frac{1}{(d-2)^{1/3}}$ it can be concluded that the state is of Schmidt rank three.

\section{Illustrations and Applications}
In this section we provide examples of qubit and qutrit mixed states as applications of our results. This paves the way for detecting states which
are useful for teleportation as well as to quantify the amount of entanglement required for teleportation, in realistic settings.

\subsection{Two qutrit mixed states with Schmidt rank two}

We consider a two qutrit mixed state with Schmidt rank two \cite{rank} given by
\begin{equation}
\rho_f=\frac{5p}{p+2} \rho_c + \frac{2(1-2p)}{p+2} |\phi\rangle \langle\phi|;   0 \le p \le \frac{1}{2},
\label{state1}
\end{equation}
where, $\rho_c=\frac{1}{2}(|\chi_0\rangle \langle\chi_0| + |\chi_1\rangle \langle\chi_1|)$. This decomposition for
state $\rho_f$ is optimal. Here, $|\chi_0\rangle$ and
$|\chi_1\rangle$ are of the form $|\chi_0\rangle = \sqrt{\frac{3}{5}} |\psi\rangle + \sqrt{\frac{2}{5}} |\phi\rangle$ and
$|\chi_1\rangle = \sqrt{\frac{3}{5}} |\psi\rangle - \sqrt{\frac{2}{5}} |\phi\rangle$, respectively, and
the states $|\psi\rangle,|\phi\rangle$ are given by,
$|\psi\rangle = \frac{1}{\sqrt{3}} (|00\rangle + |11\rangle - e^{\frac{i\pi}{3}} |22\rangle)$ and
$|\phi\rangle = \frac{1}{\sqrt{2}} (|00\rangle + |11\rangle)$. Also, $p$ is the classical probability of mixing.

We check whether the bounds on
$E^{(3,2)}_2(\rho_i)$ works  for the above density matrix. For $3\otimes 3$ dimension, $E^{(3,2)}_2 (\rho_f)$  (see 
Eq. (\ref{new45})) becomes
\begin{eqnarray}
 E^{(3,2)}_2 (\rho_f)&=&\frac{3\sqrt{3}}{2}\left(\min\displaystyle\sum_{i} p_if_2(\rho_i)\right)-\frac{\sqrt{3}}{2}\nonumber\\
&=&\frac{3\sqrt{3}}{2}\left(\min_{\{p\}}\left[\frac{1+p}{2+p}\right]\right)-\frac{\sqrt{3}}{2}\nonumber\\
&=&\frac{\sqrt{3}}{4}; \hspace{0.5cm}\mbox{for $p=0$}.
\label{32}
\end{eqnarray}
In this calculation we have used the appropriate maximally 
entangled basis given in \cite{Karim}.
From Eqs. (\ref{32}) and (\ref{entschmidt2}),
it can be seen that the state (in Eq. (\ref{state1})) is useful for teleportation.

\subsection{States generated as a result of Two-Qubit Interaction with a Squeezed Thermal Bath}

Open quantum systems is the systematic study of the evolution of the system of interest, such as a qubit, under the influence of
its environment, also called the bath or the reservoir. This results in decoherence and dissipation.
Consider the Hamiltonian $H=H_S+H_R+H_{SR}$; where
$S$ stands for the system of interest, $R$ for reservoir and $SR$ for the system-reservoir interaction. Depending upon the system-reservoir
interaction,  open systems can be classified into two broad categories, viz., dissipative or QND (quantum non-demolition). In case
of QND dephasing occurs without damping the system, i.e., where $[H_S,H_{SR}]=0$ while decoherence along with dissipation
occurs in dissipative systems, i.e., $[H_S,H_{SR}]\neq 0$. \cite{sbrg, sbgp, sbsq}.

\subsubsection{States generated as a result of Two-Qubit Open System Interacting with a Squeezed Thermal Bath via a Dissipative Interaction}

 Here we study the dynamics of the bound [Eq. (\ref{BE32})] for a two-qubit open system interacting with a
squeezed thermal bath, modeled as a $3-D$ electromagnetic field (EMF),  as well as its specialization to a vacuum bath, where the bath squeezing ($r$) and temperature
($T$) are set to zero, and undergoing a dissipative interaction \cite{sb10}.
The model Hamiltonian is
\begin{eqnarray}
 H&=&H_S+H_R+H_{SR}\nonumber\\
 &=&\displaystyle\sum_{n=1}^2\hbar\omega_nS_n^z+\displaystyle\sum_{\vec{k}_s}\hbar\omega_k\left(b_{\vec{k}_s}^\dagger
 b_{\vec{k}_s}+\frac{1}{2}\right)\nonumber\\ {}&-&i\hbar\displaystyle\sum_{\vec{k}_s}\displaystyle\sum_{n=1}^2[\vec{\mu}_n.
 \vec{g}_{\vec{k}_s}(\vec{r}_n)(S_n^++S_n^-)b_{\vec{k}_s}-h.c.].
\end{eqnarray}
Here $\vec{\mu}_n$ are the transition dipole moments, dependent on the different atomic positions $\vec{r}_n$ and $S_{n}^{+} $ $(=
\frac{1}{2}|e_n\rangle\langle g_n|)$, and $S_n^-(=\frac{1}{2}|g_n\rangle \langle e_n|)$ are the dipole raising and
lowering operators satisfying the usual commutation relations. $S_n^z(=\frac{1}{2}(|e_n\rangle\langle e_n|
-|g_n\rangle\langle g_n|))$ is the energy operator of $n$th atom and $b_{\vec{k}_s}^\dagger$, $b_{\vec{k}_s}$ are the creation
and anihilation operators of the field mode $\vec{k}_s$ with the wave vector $\vec{k}$ and polarization index $s=1,2$. A key feature of the model is that the
system-reservoir (S-R) coupling constant $\vec{g}_{\vec{k}_s}(\vec{r}_n)$ is dependent on the position of the qubit $r_n$ and is
\begin{equation}
 \vec{g}_{\vec{k}_s}(\vec{r}_n)=\left(\frac{\omega_k}{2\epsilon\hbar V}\right)^{\frac{1}{2}}\vec{e}_{\vec{k}_s}e^{i\vec{k}.r_n},
\end{equation}
where $V$ is the normalization volume and $\vec{e}_{\vec{k}_s}$ is the unit polarization vector of the field. The position
dependence of the coupling leads to interesting dynamical consequences and allows the entire dynamics to be classified into
two categories, that is, the independent regime, where the interqubit distance is far enough for each qubit to locally interact with
an independent bath or the collective regime, where the qubits are close enough for them to interact with the bath collectively.
Asuming  an initial system-reservoir separable state, with the system in a separable, and the bath in a
squeezed thermal state, with time the qubits develop correlations between themselves via a channel setup by the bath. A master equation
for the reduced dynamics of the two qubit system is obtained by tracing out the environment
(bath), using the usual Born-Markov and rotating wave approximation (RWA). This can be then solved to obtain the dynamics of the reduced density matrix, whose
details are presented in \cite{sb10}, for the general case of a squeezed thermal bath at finite
temperature as well as for a vacuum reservoir.

Let the reduced
two-qubit density matrix of the system be $\rho_f(t)$. Its
spectral decomposition corresponding to its eigenvalues
($\lambda_i(t) $) is,
\begin{equation}
 \rho_f(t)=\displaystyle\sum_i \lambda_i(t) \rho_i(t).
\end{equation}
Here $\rho_i(t)=|\psi_i(t)\rangle\langle\psi_i(t)|$, $|\psi_i(t)\rangle$ being the eigenvectors corresponding
to the eigenvalues $\lambda_i(t) $ ($\sum_i \lambda_i (t) = 1$). For a two qubit state the
Eq. (\ref{BE32}) becomes
\begin{equation}
 0 \leq E^{(2,2)}_2 (\rho_f(t)) \leq  1.
\label{22}
\end{equation}
We can easily say that for two-qubit state $E^{(2,2)}_2$ is
nothing but concurrence $C$. If we look at the Figs.
(\ref{con1}), and (\ref{con2}), for the case of  a vacuum bath ($T=0, r=0$),
concurrence $C$ (or $E^{2,2}_2$) is
seen to decrease with time of evolution $t$,
with a predominantly oscillatory behavior in the collective regime (marked by the inter qubit distance
$r_{12}<1$). The singlet
fraction $f$ also shows similar behavior. From these two figures,
it is clear that when and where $C$  becomes zero, and $f$ is equal to $\frac{1}{2}$.

For the case of a squeezed thermal bath, as the system evolves with time $t$, concurrence $C$ and $f$
exhibit damped behavior, as seen in Figs. (\ref{con3}) and (\ref{con4}). If  we increase the
inter-qubit distance $r_{12}$, then the concurrence $C$ for the system suddenly falls to zero (i.e., sudden death of
entanglement in the system). Thus, the system can be used as a resource for teleportation purpose in the range
$0\le r_{12} < r_d$. Here we define a new term $r_d$, such that at $r_{12}=r_d$ concurrence $C$ of the system becomes zero.
Obviously this $r_d$ will be different for different parameter ($T, r$) settings. The Figs.
(\ref{con4}) depict the abrupt decrease of  concurrence $C$ and singlet fraction $f$ as $r_{12}$ increases.
In Figs. (\ref{con7}), the behavior of $C$ and $f$ with respect to environmental squeezing parameter $r$ is shown.
For $r$ between $-0.02$ to $0.02$ both $C$ and $f$ remain almost constant, thereby exhibiting the tendency of squeezing
to resist environmental degradation. Beyond this range there is a rapid fall of the depicted quantities.
\begin{widetext}
\begin{center}
\begin{figure}
\[
\begin{array}{cc}
\includegraphics[height=4cm,width=5.5cm]{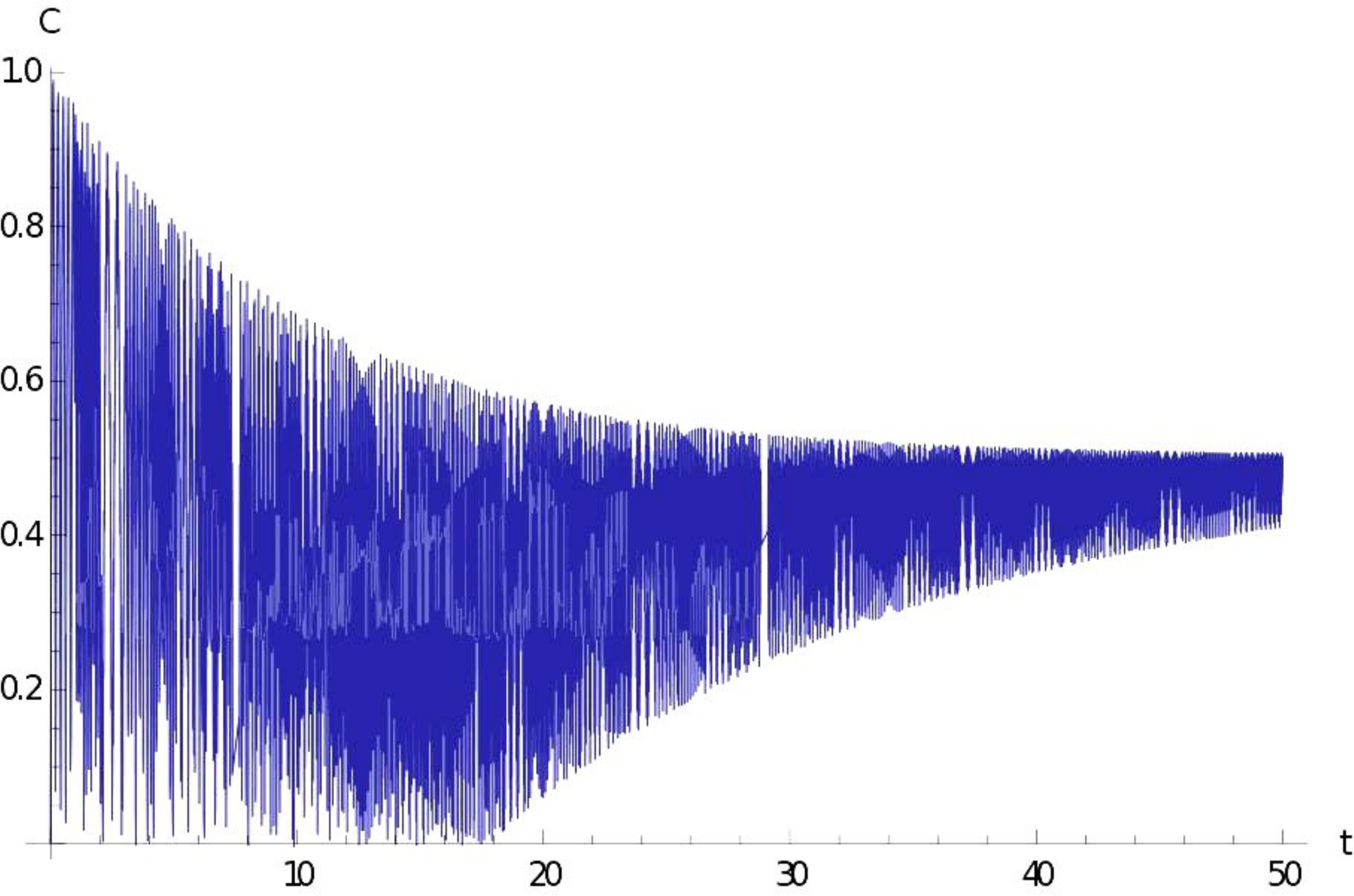}&
\includegraphics[height=4cm,width=5.5cm]{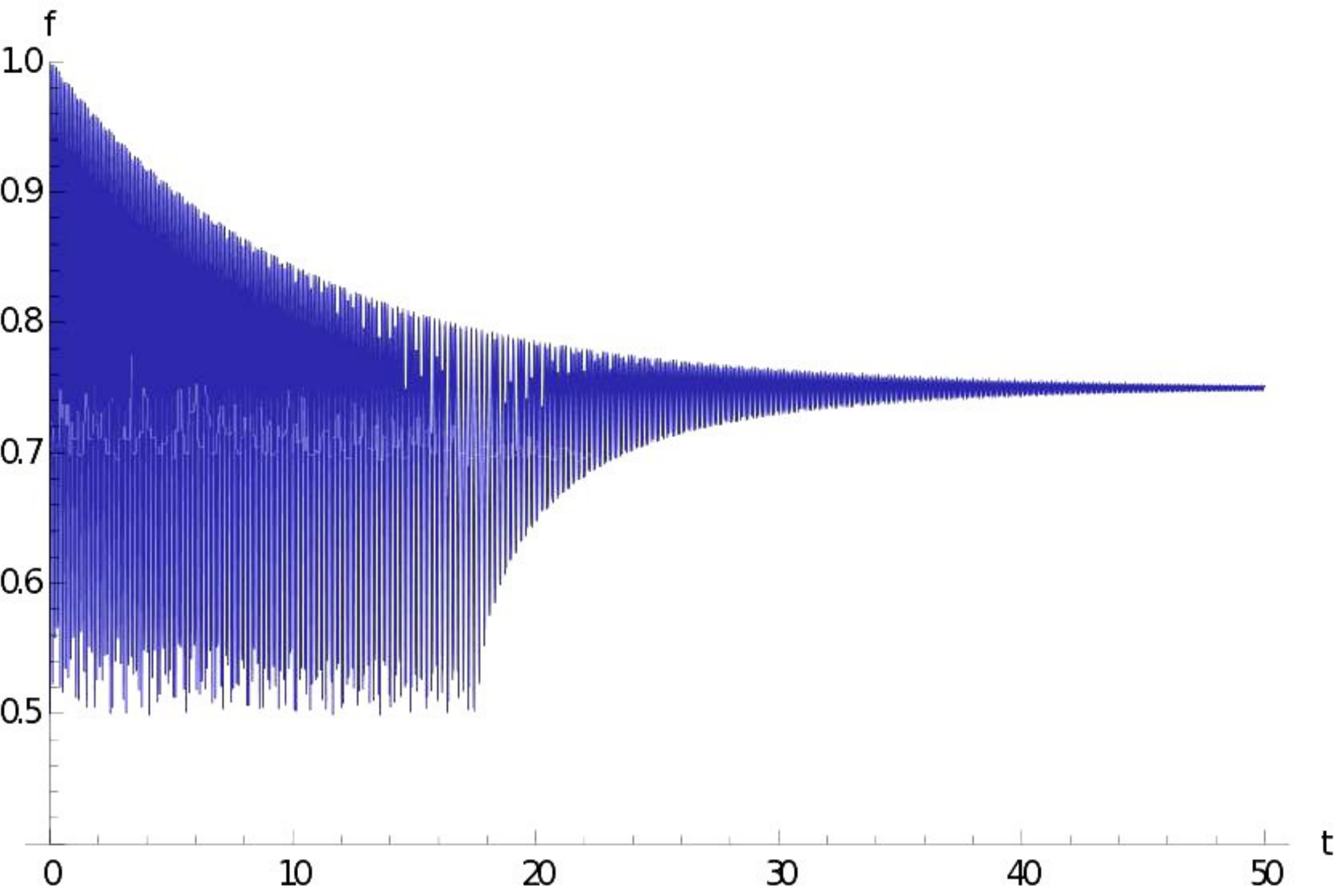}\\
(i)&(ii)
\end{array}
\]
\caption{Plot of (i) concurrence $C$ (or $E^{(2,2)}_2$) and (ii) singlet fraction $f$
with respect to the time of evolution $t$, respectively. Here we consider the
case of a vacuum bath ($T=r=0$) and the collective decoherence model ($r_{12}=0.05$).} \label{con1}

\end{figure}
\begin{figure}

\[
\begin{array}{cc}
\includegraphics[height=4cm,width=5.5cm]{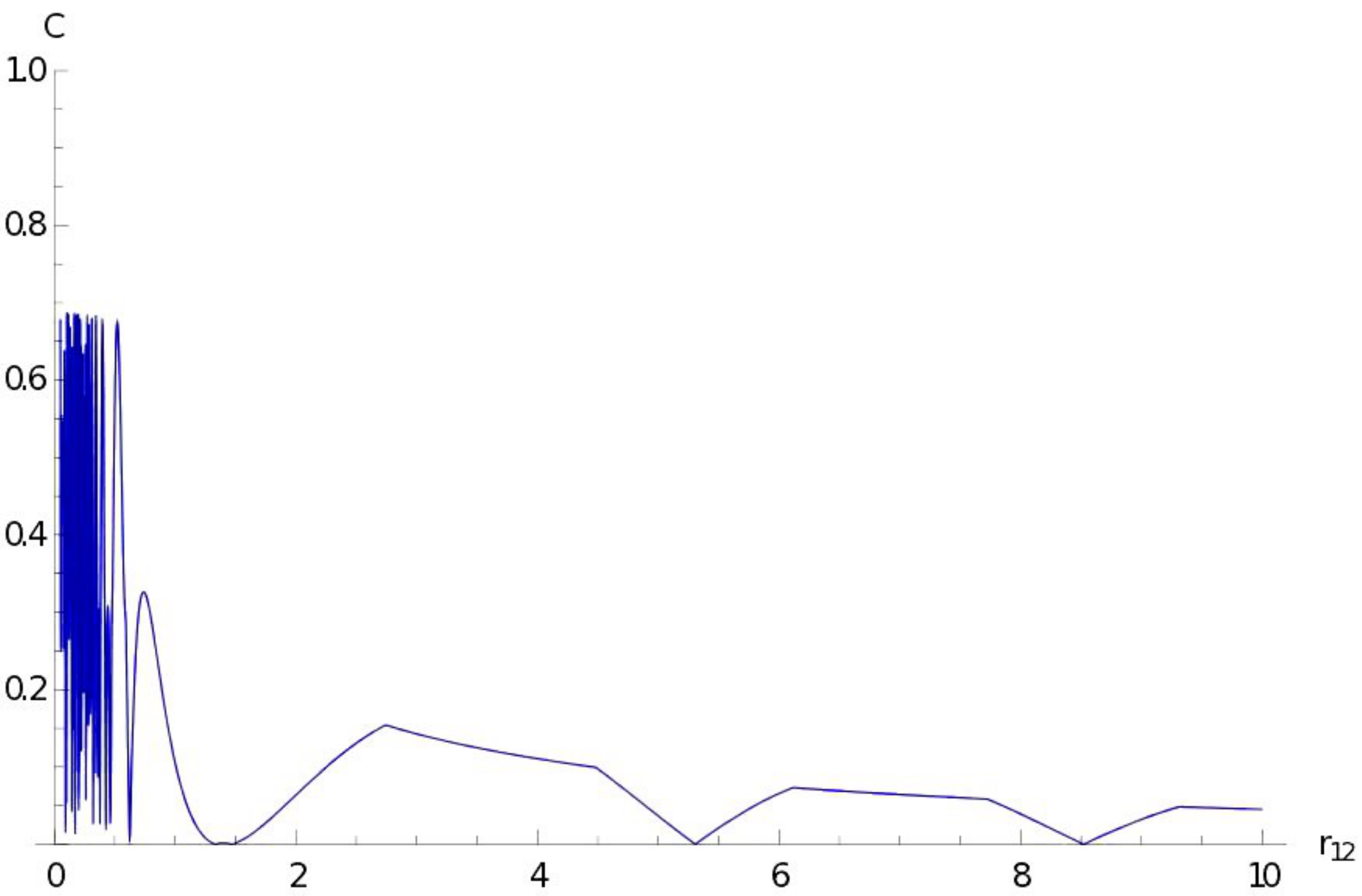}&
\includegraphics[height=4cm,width=5.5cm]{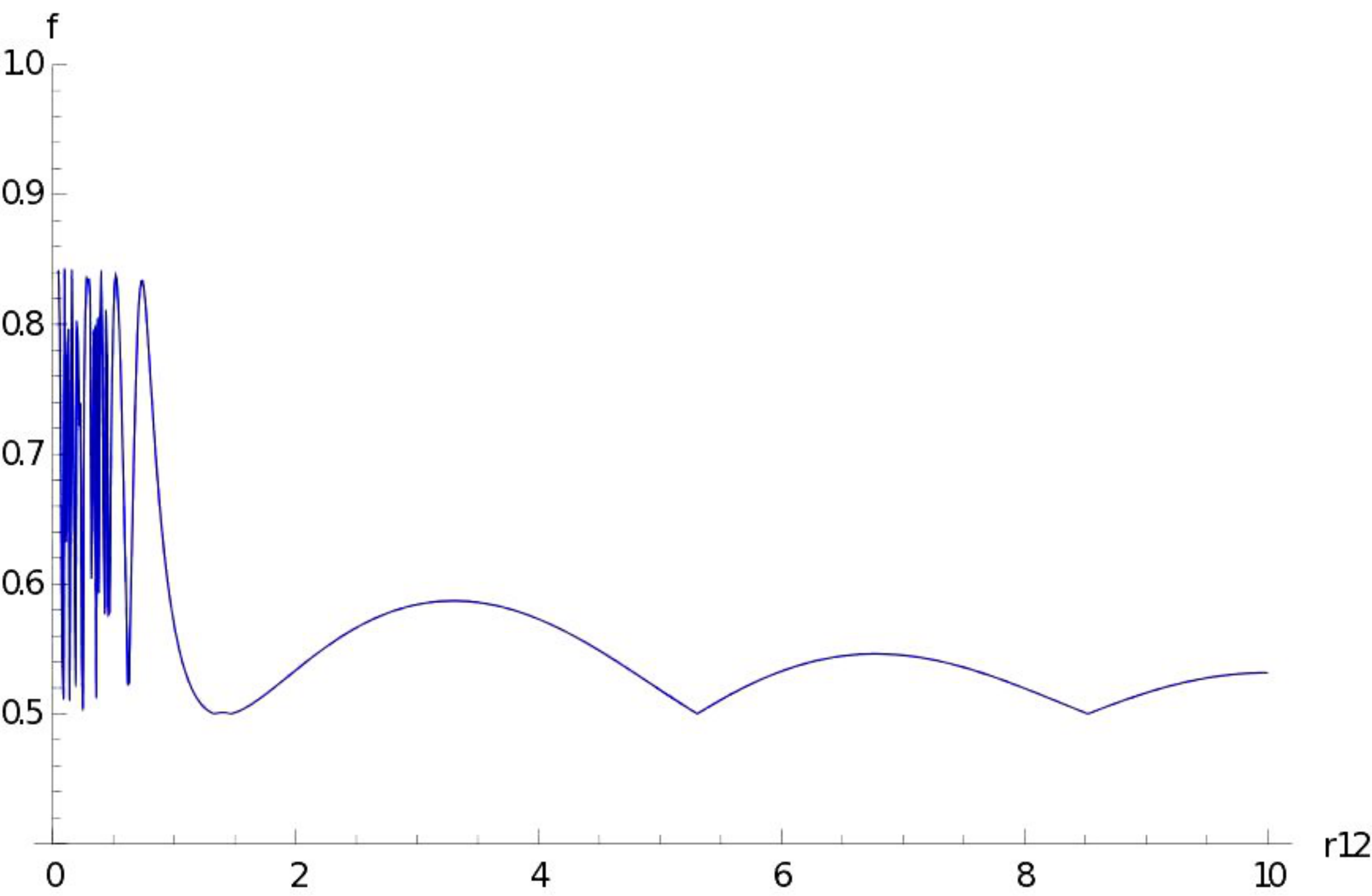}\\
(i)&(ii)
\end{array}
\]
\caption{Plot of (i) concurrence $C$ (or $E^{(2,2)}_2$) and (ii) $f$
with respect to the inter-qubit distance $r_{12}$, respectively. Here we consider the
case of vacuum bath ($T=r=0$) and system is at time $t=10$.} \label{con2}
\end{figure}
\begin{figure}

\[
\begin{array}{cc}
\includegraphics[height=4cm,width=5.5cm]{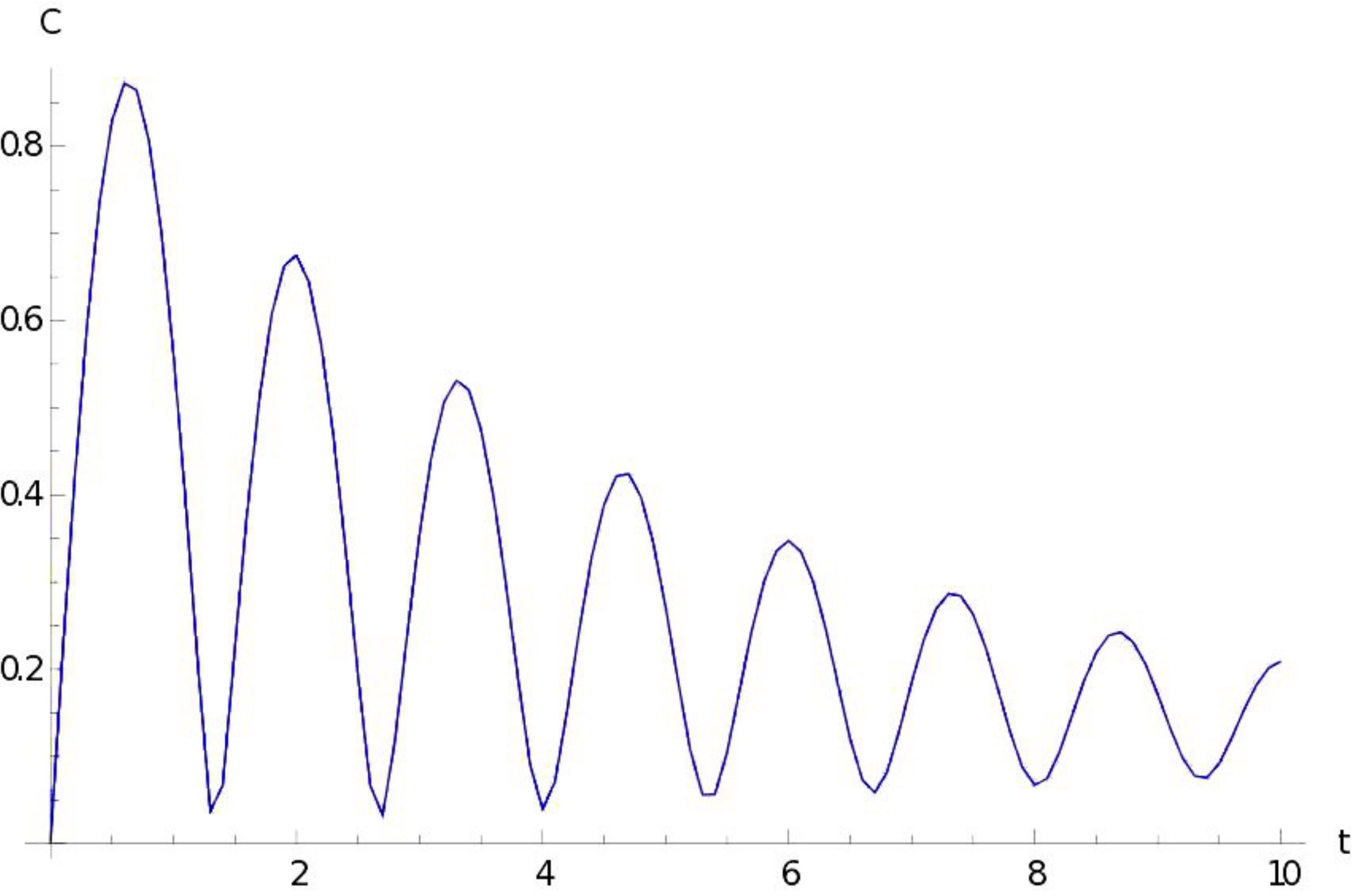}&
\includegraphics[height=4cm,width=5.5cm]{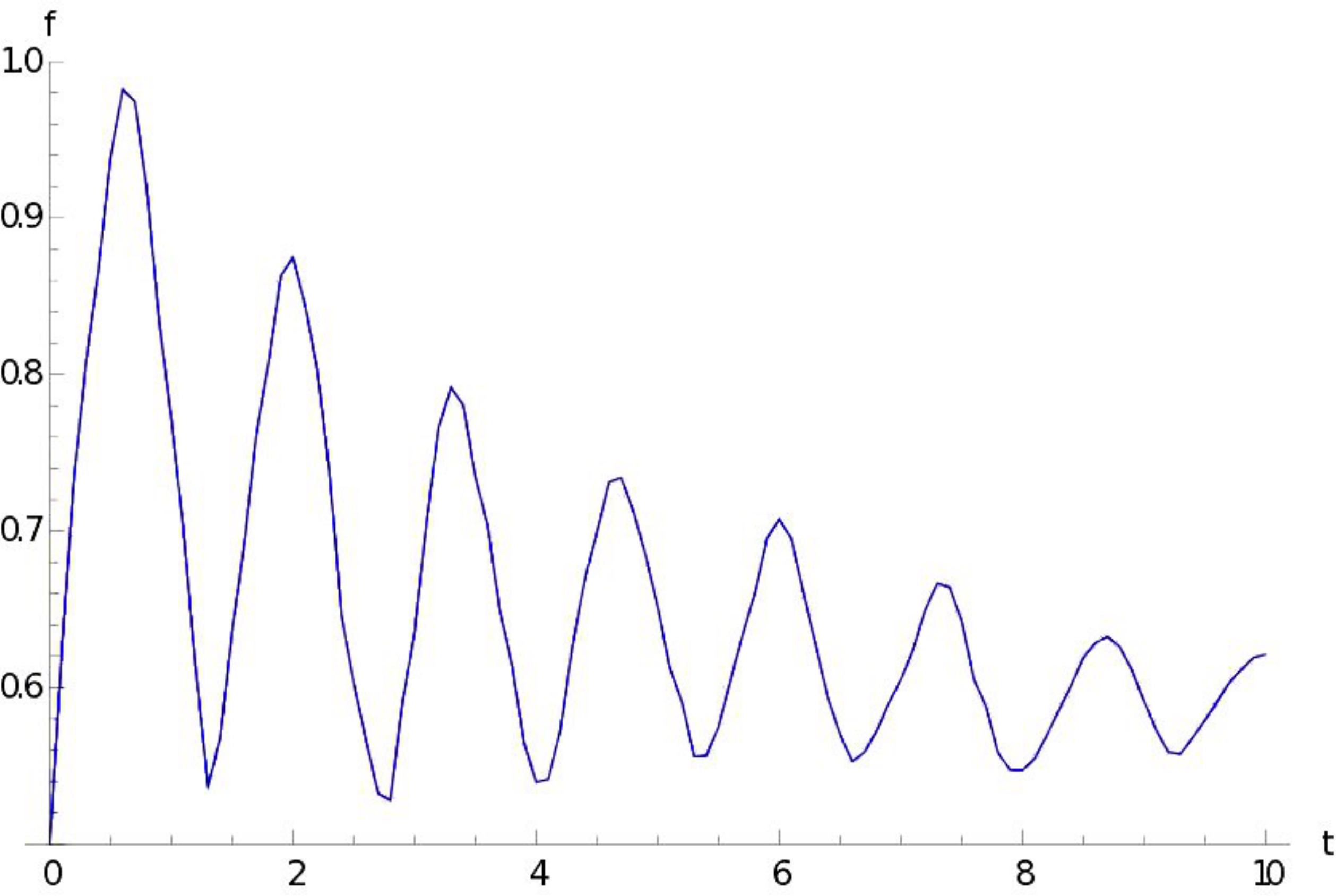}\\
(i)&(ii)
\end{array}
\]
\caption{ Plot of (i) concurrence $C$ (or $E^{(2,2)}_2$) and (ii) singlet fraction $f$
with respect to the time of evolution $t$, respectively, for a squeezed thermal bath ($T=1, r=0.1$) in the collective regime ($r_{12}=0.05$).} \label{con3}
\end{figure}
\begin{figure}

\[
\begin{array}{cc}
\includegraphics[height=4cm,width=5.5cm]{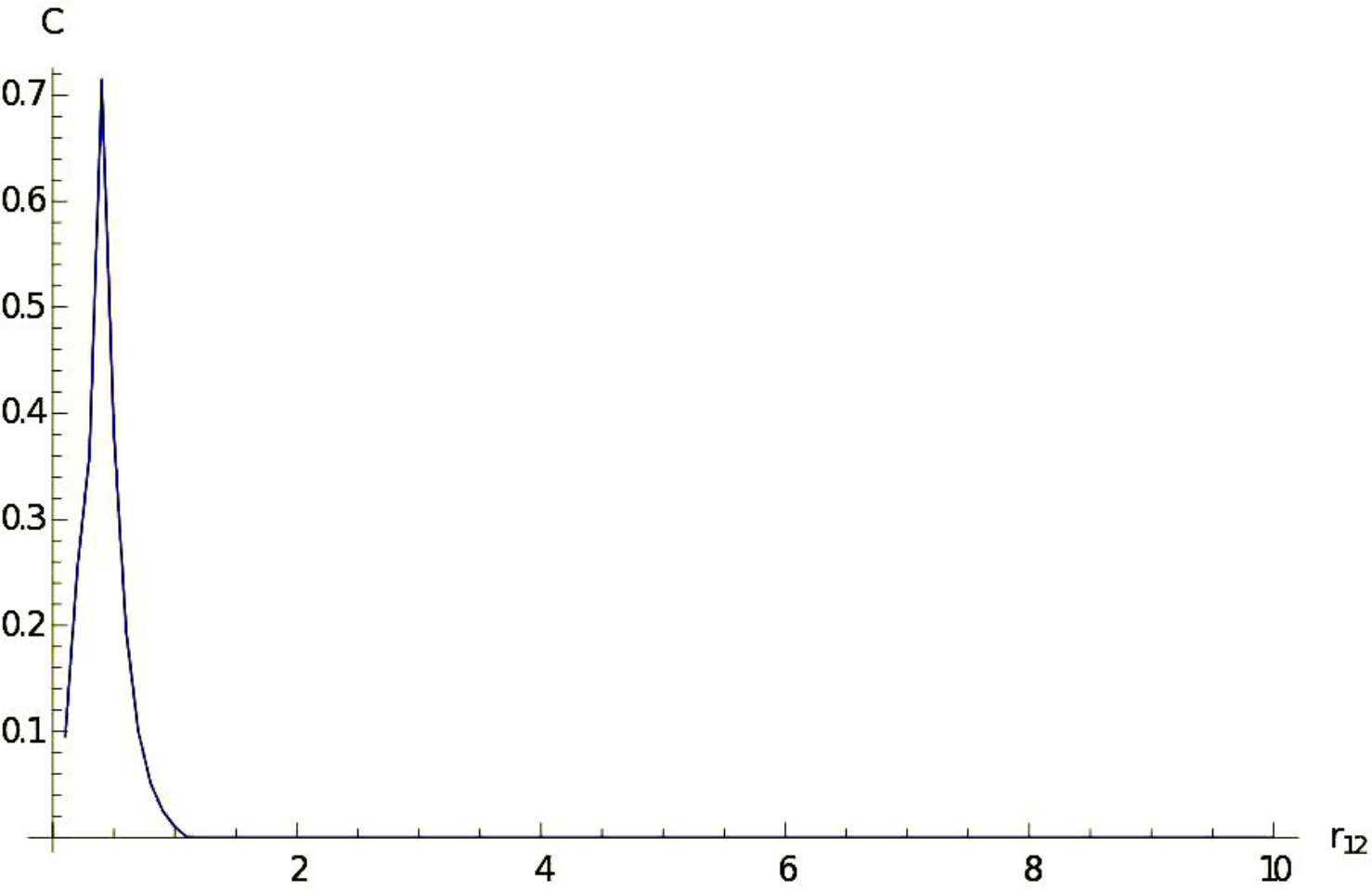}&
\includegraphics[height=4cm,width=5.5cm]{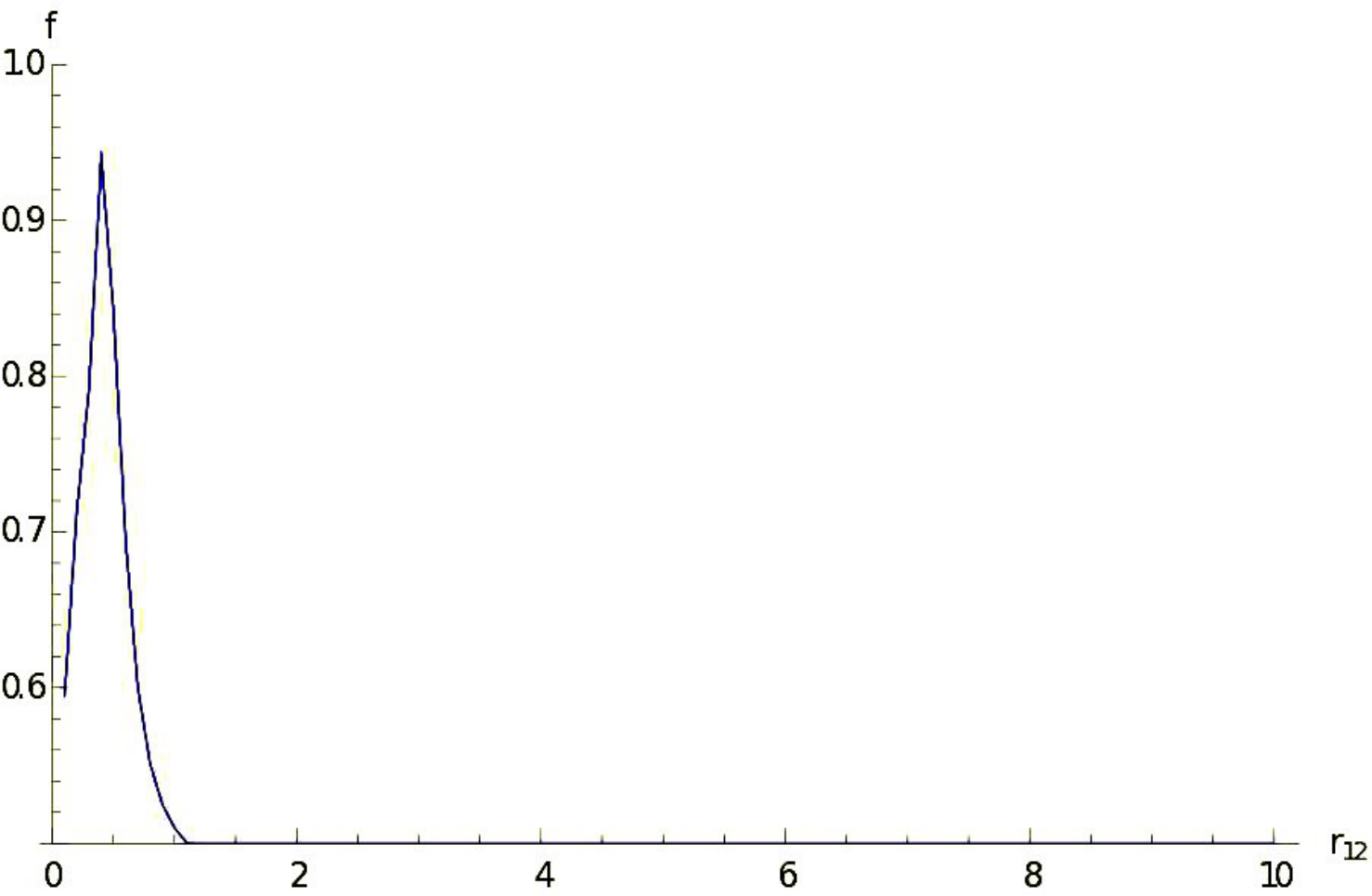}\\
(i)&(ii)
\end{array}
\]
\caption{Plot of (i) concurrence $C$ (or $E^{(2,2)}_2$) and (ii) $f$
with respect to the inter-qubit distance $r_{12}$, respectively, for a squeezed thermal bath ($T=1, r=0.1$) and 
time of evolution $t=1$.} \label{con4}
\end{figure}
\begin{figure}

\[
\begin{array}{cc}
\includegraphics[height=4cm,width=5.5cm]{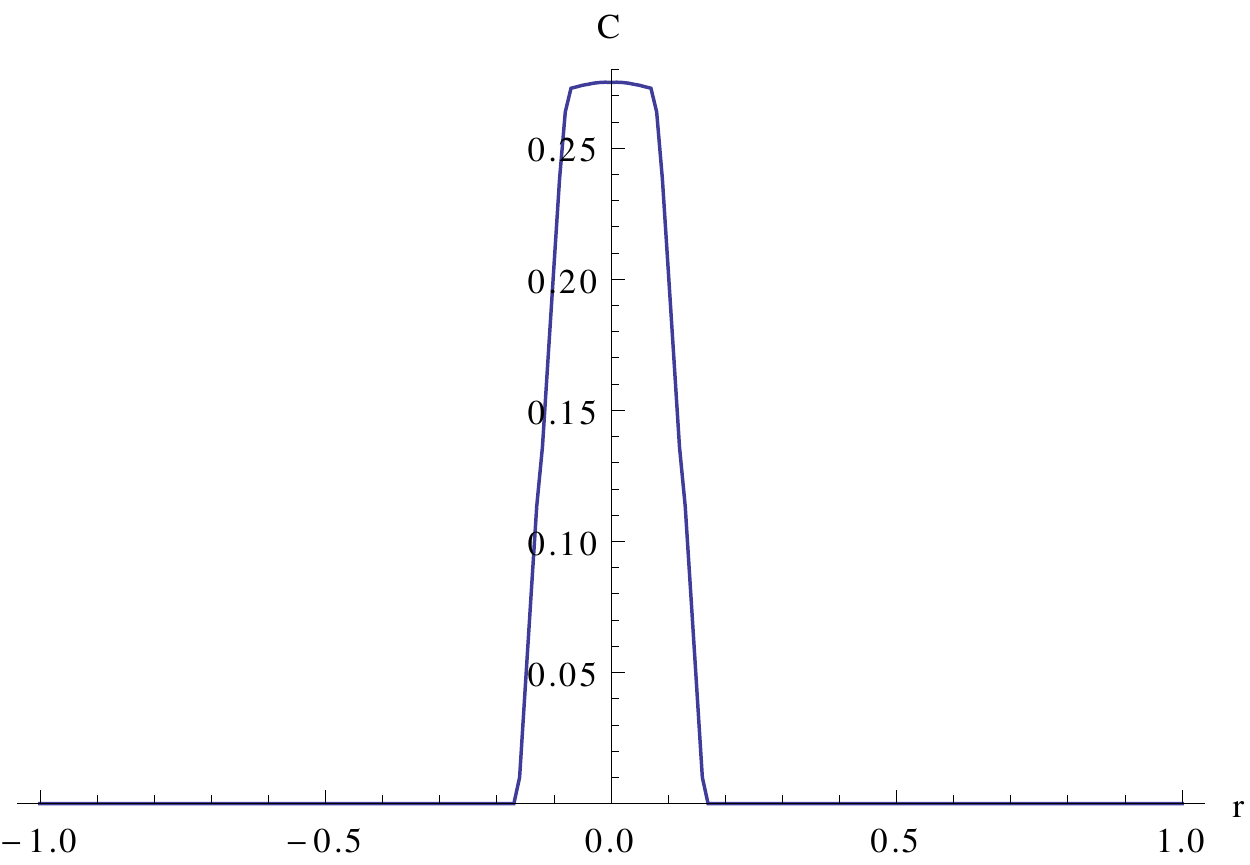}&
\includegraphics[height=4cm,width=5.5cm]{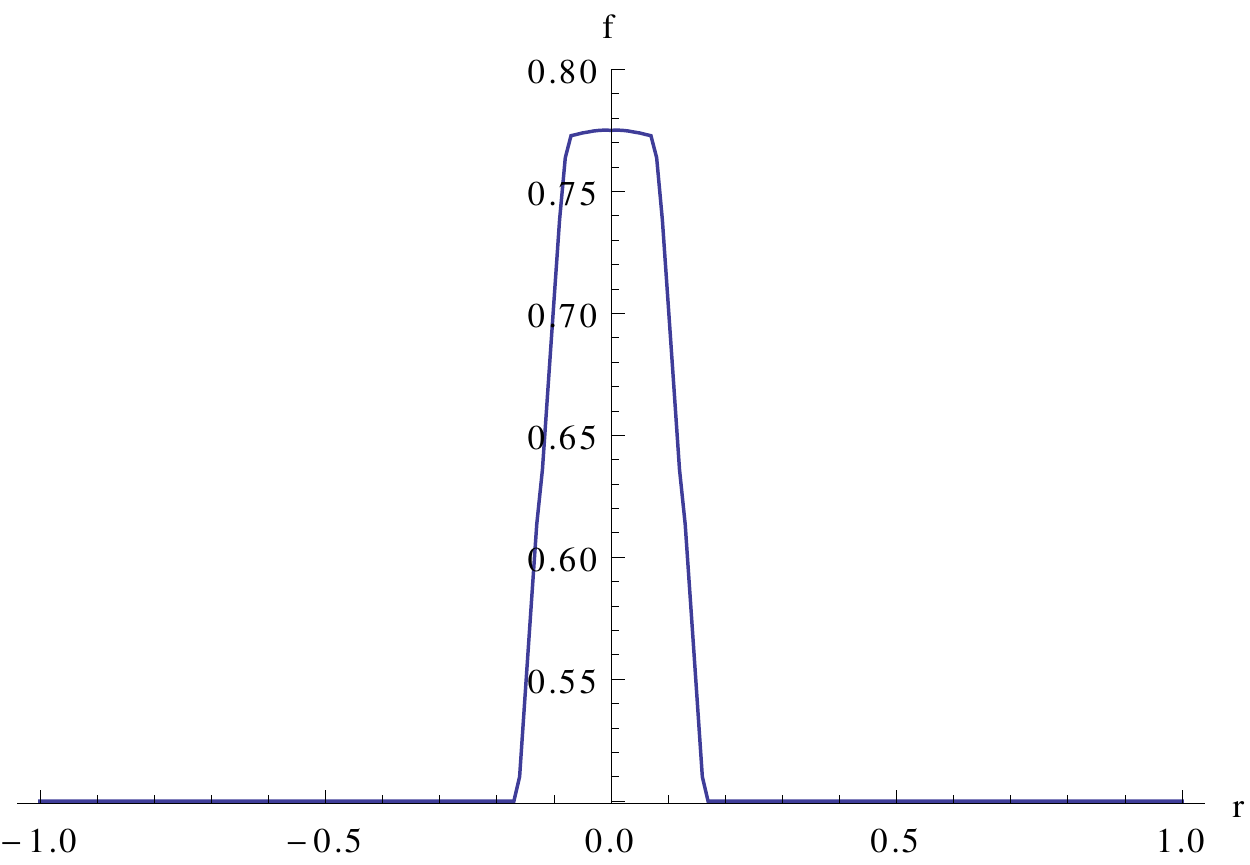}\\
(i)&(ii)
\end{array}
\]
\caption{Plot of (i) concurrence $C$ (or $E^{(2,2)}_2$) and (ii) $f$
with respect to the squeezing parameter $r$, respectively, for a thermal bath ($T=5, r_{12}=0.05$) and 
time of evolution $t=2$.} \label{con7}
\end{figure}
\begin{figure}

\[
\begin{array}{cc}
\includegraphics[height=4cm,width=5.5cm]{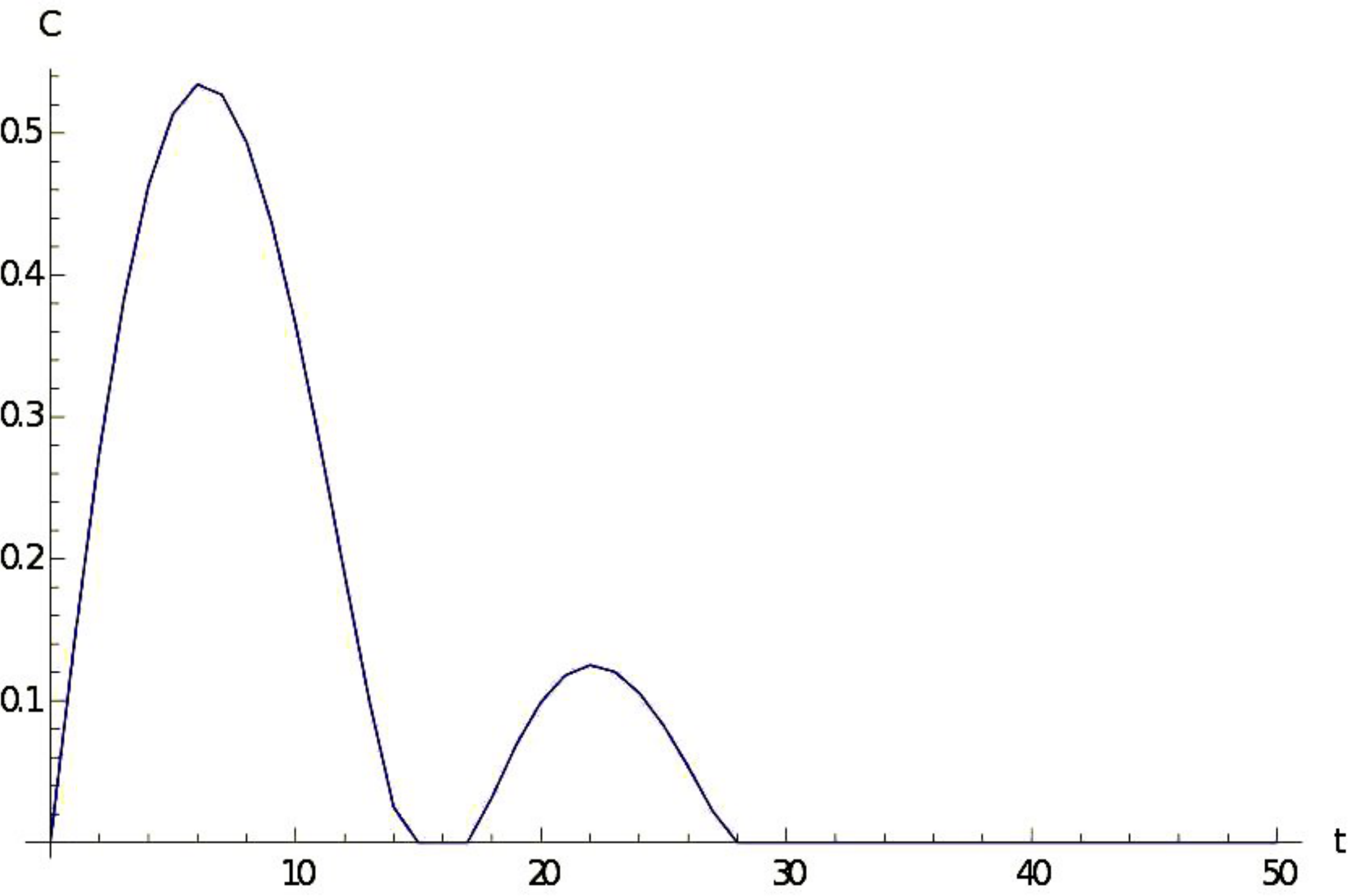}&
\includegraphics[height=4cm,width=5.5cm]{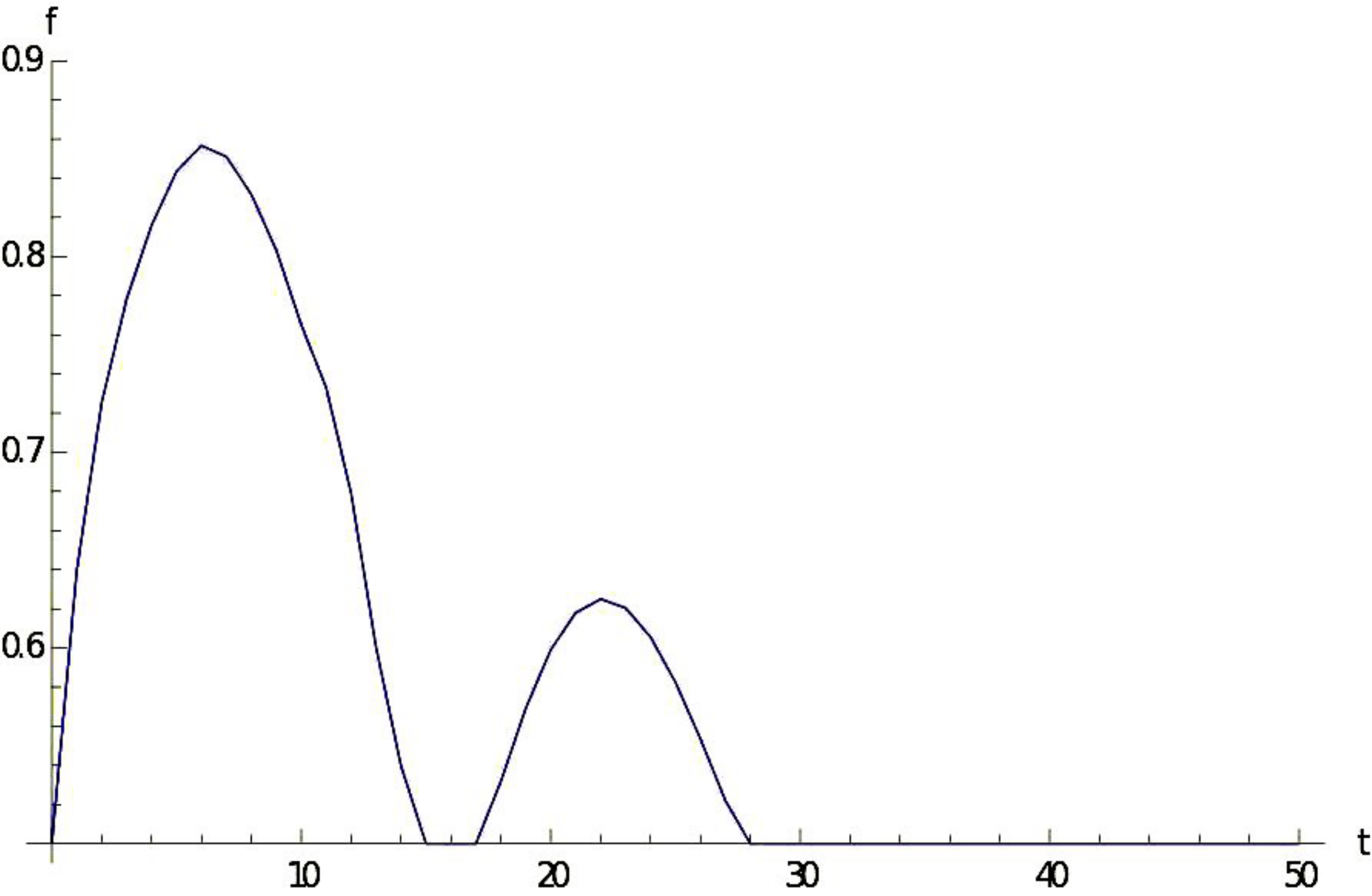}\\
(i)&(ii)
\end{array}
\]
\caption{Plot of (i) concurrence $C$ (or $E^{(2,2)}_2$)  and (ii) $f$
as a function of the time of evolution $t$. Here we consider the
case of QND interaction ($T=5, r=0.1$), in the collective decoherence regime ($r_{12}=0.05$).} \label{con5}
\end{figure}
\begin{figure}

\[
\begin{array}{cc}
\includegraphics[height=4cm,width=5.5cm]{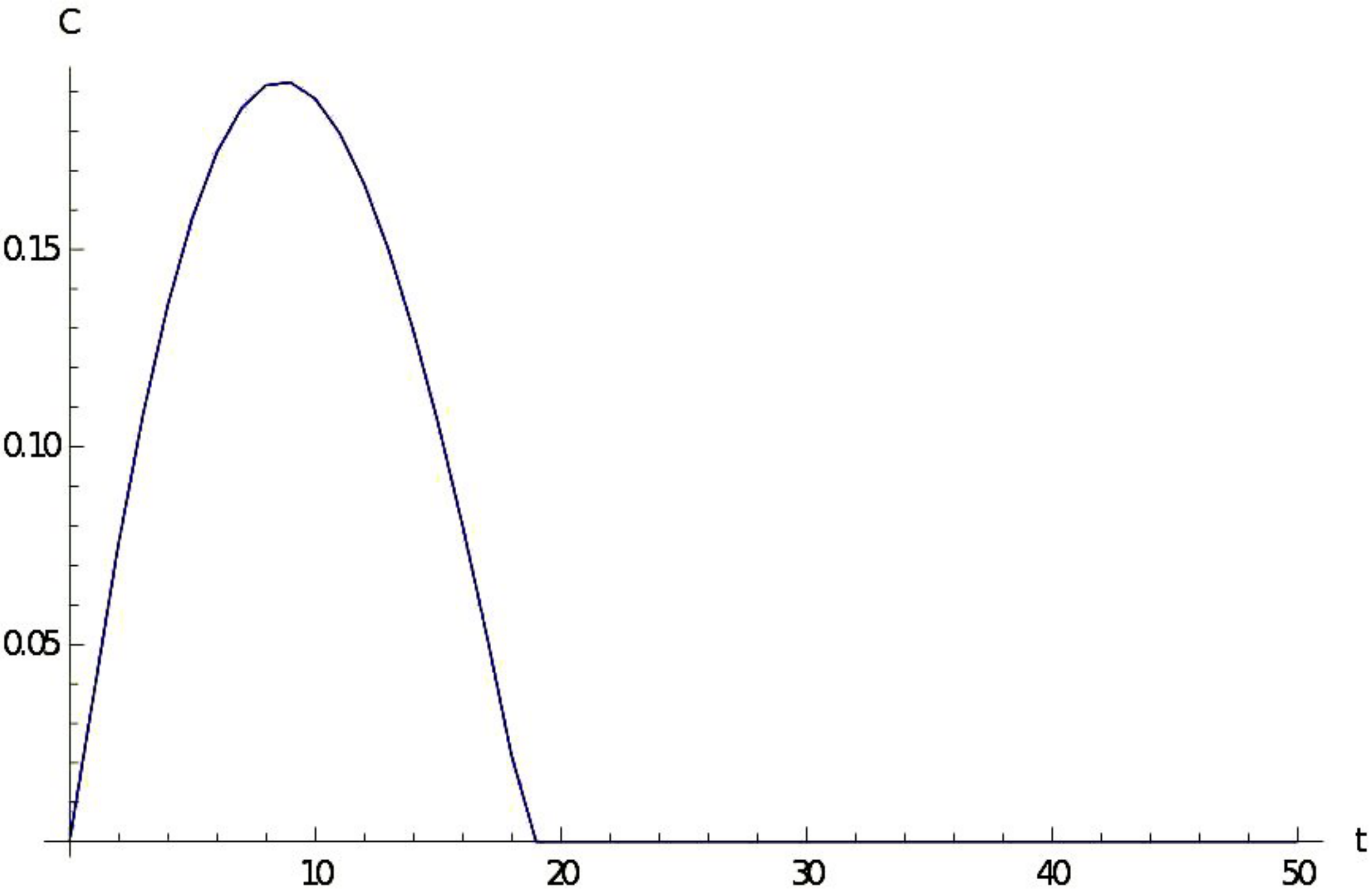}&
\includegraphics[height=4cm,width=5.5cm]{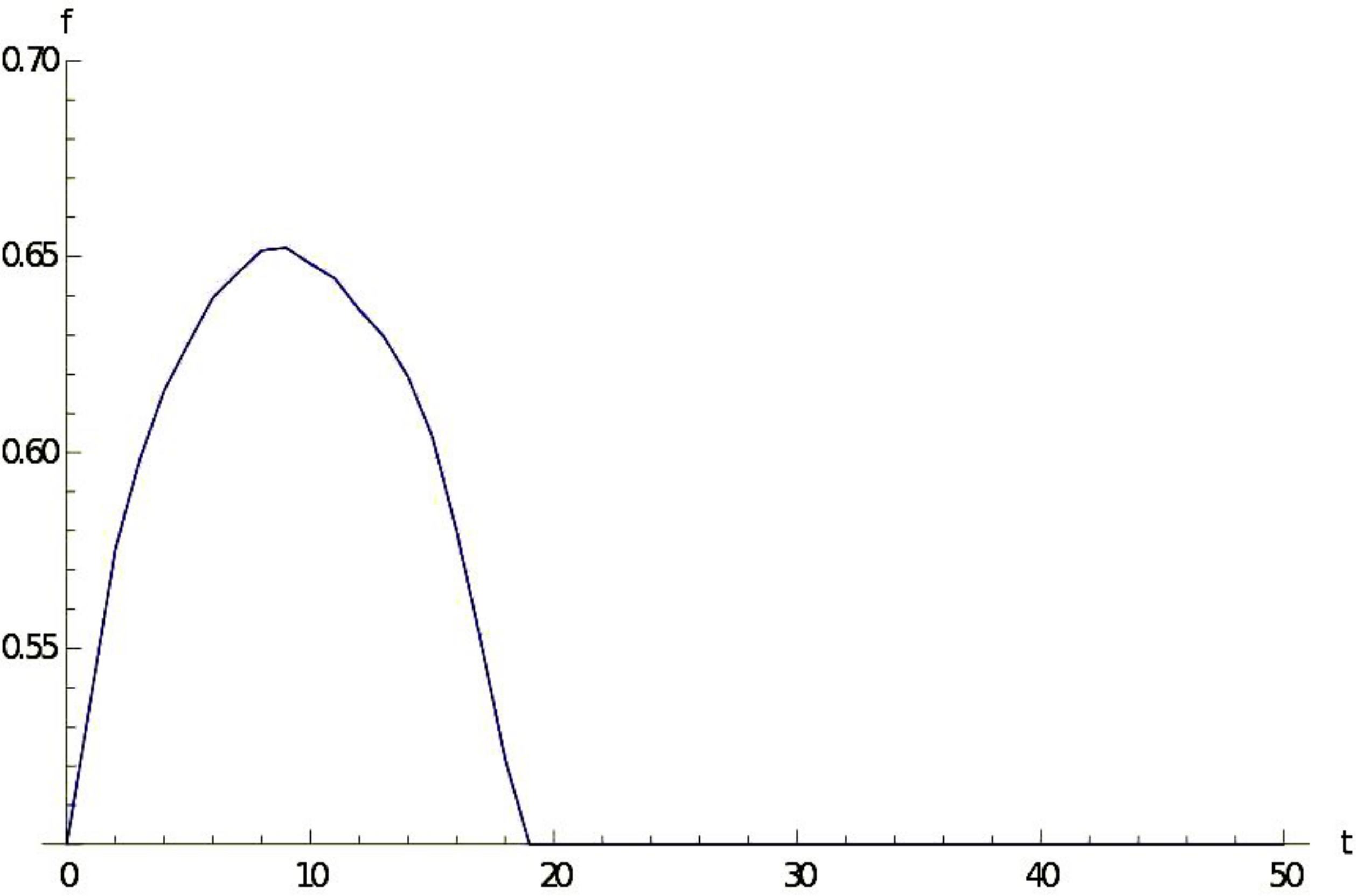}\\
(i)&(ii)
\end{array}
\]
\caption{Plot of (i) concurrence $C$ (or $E^{(2,2)}_2$) and (ii) $f$
as a function of the time of evolution $t$, for the 
case of QND interaction ($T=5, r=0.1$), in the independent decoherence regime ($r_{12}=1.1$).} \label{con6}
\end{figure}
\begin{figure}

\[
\begin{array}{cc}
\includegraphics[height=4cm,width=5.5cm]{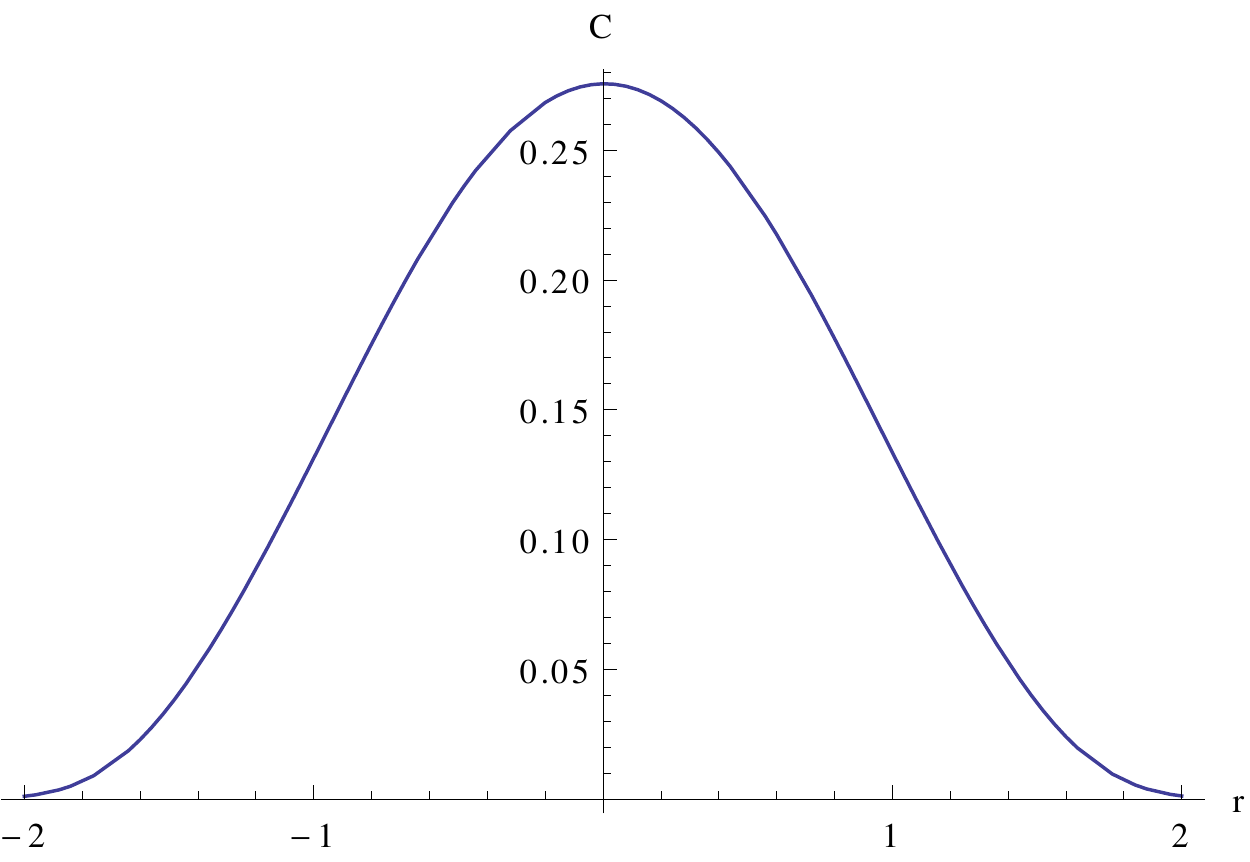}&
\includegraphics[height=4cm,width=5.5cm]{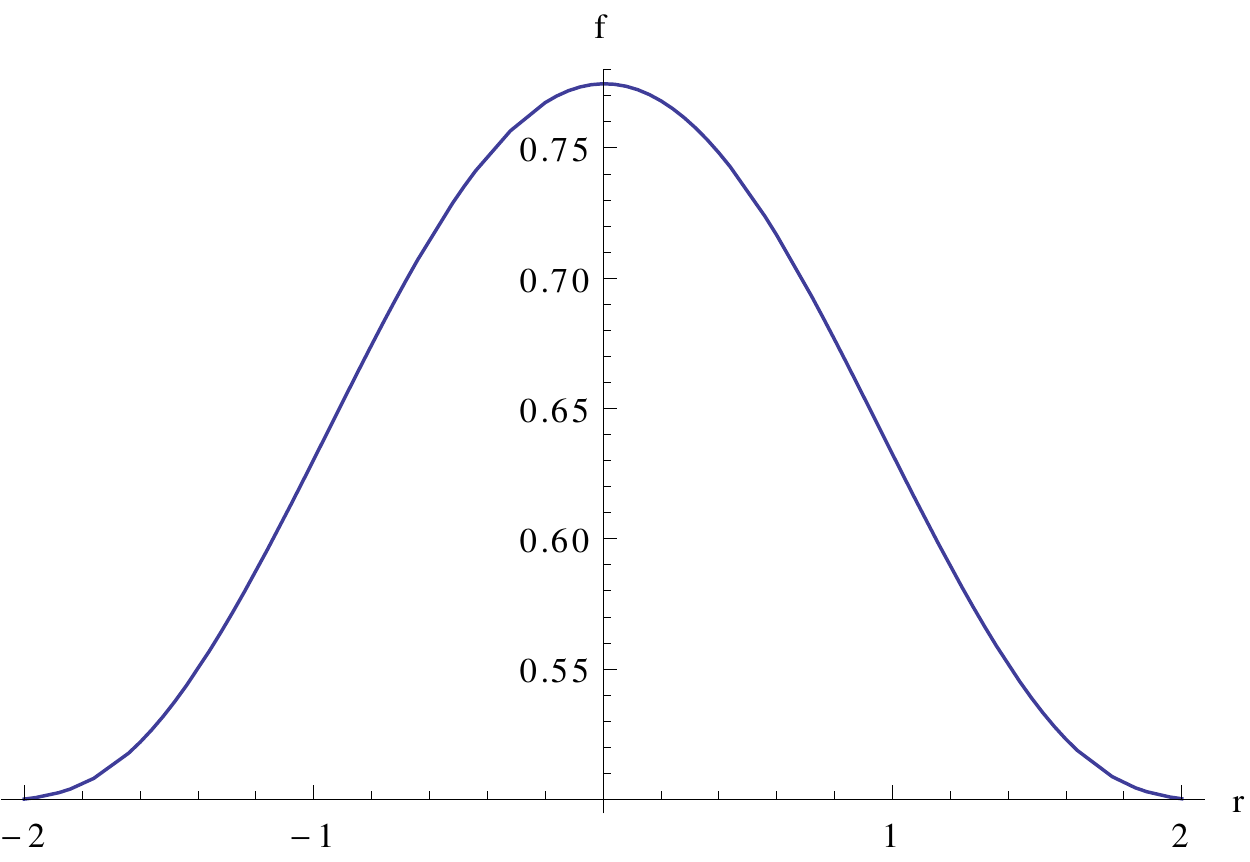}\\
(i)&(ii)
\end{array}
\]
\caption{Plot of (i) concurrence $C$ (or $E^{(2,2)}_2$)  and (ii) $f$
as a function of squeezing parameter $r$. Here we consider the
case of QND interaction ($T=5$), in the collective decoherence regime ($r_{12}=0.05$) and 
time of evolution $t=2$.} \label{con8}
\end{figure}
\end{center}
\end{widetext}

\subsubsection{States generated as a result of Two-Qubit Open System Interacting with a Squeezed Thermal Bath via Quantum
Nondemolition Interaction}

Now we  take up the Hamiltonian,  describing a  QND interaction  of two
qubits with the bath as
\begin{eqnarray}
H & = & H_S + H_R + H_{SR} \nonumber \\ & = & \sum\limits_{n=1}^2 \hbar
\varepsilon_n J^n_z+ \sum\limits_k \hbar \omega_k b^{\dagger}_k
b_k \nonumber\\  &+&  \sum\limits_{n,k} \hbar J^n_z (g^n_k b^{\dagger}_k +
g^{n*}_k b_k). \label{basich}
\end{eqnarray}
Here  $H_S$, $H_R$  and $H_{SR}$  stand  for the  Hamiltonians of  the
system,  reservoir  and  system-reservoir  interaction,  respectively.
$b^{\dagger}_k$, $b_k$ denote  the creation and annihilation operators
for the  reservoir oscillator of frequency  $\omega_k$, $g^n_k$ stands
for the coupling  constant (again assumed to be position  dependent) for the
interaction  of the  oscillator  field with  the  qubit system and are
taken to be
$g^n_k = g_k e^{-ik.r_n}$, 
where  $r_n$ is  the  qubit position.   Since  $[H_S, H_{SR}]=0$,  the
Hamiltonian [Eq. \ref{basich})] is of QND type. In the parlance of quantum information
theory,  the  noise  generated  is  called  the  phase  damping  noise. The position
dependence of the coupling constant once more allows for the dynamical classification
into the independent and collective regimes. In order to obtain the reduced  dynamics of the system , we trace over
the reservoir  variables, the details of which can be found in \cite{sbqnd}.

Now we study the behavior of concurrence $C$ (actually $E^{2,2}_2$) and singlet fraction $f$ as the two-qubit
system evolves with time $t$ both for collective and localized (independent) decoherence model.
It can be noticed from  Figs. (\ref{con5}), and (\ref{con6}) that the value of
concurrence $C$ is higher and lasts longer in the case of collective decoherence model than in the case of localized
decoherence model. As expected, the the singlet fraction $f$ shows similar kind
of behavior with time $t$. When $C$ becomes zero, $f$ becomes equal to $\frac{1}{2}$, i.e., the system at
this particular time $t$ cannot be useful for teleportation, otherwise it is useful. Hence
the system satisfies the lower bound of Eq. (\ref{22}), when concurrence $C$ vanishes.
The behavior of $C$ and $f$, under pure dephasing, with respect to environmental squeezing parameter $r$ is depicted in Figs. (\ref{con8}).
For $r$ between $-0.02$ to $0.02$ both $C$ and $f$ remain almost constant, thereby exhibiting the tendency of squeezing
to resist environmental degradation. Beyond this range there is a fall of the depicted quantities, though the degradation here is smoother than that in
Figs. (\ref{con7}).

\section{Conclusion}

We have made a study of entanglement of teleportation for arbitrary $d \otimes d$ dimensional states having Schmidt rank upto three.
We found that there is a simple relation between
negativity and teleportation fidelity for pure states but for mixed states, an upper bound was obtained for negativity in terms
of teleportation fidelity using convex-roof extension negativity (CREN).  The existence of a strong conjecture in the literature concerning all PPT
entangled states, in $3 \times 3$ systems, having Schmidt rank two, motivated us to develop measures capable of identifying states useful for teleportation and dependent
on the Schmidt number. This enabled a classification of entanglement as a function of teleportation fidelity, the ``Entanglement of Teleportation''. 
These results were then extended to mixed two qudit states, which we illustrated on specific examples of a two qutrit mixed state with Schmidt rank two, and two qubit states dynamically 
generated by interaction with an appropriate reservoir, for both pure dephasing as well as 
dissipative interactions. This work thus brings into focus the utility of studying higher dimensional entangled states 
using measures like ``Entanglement of Teleportation'' along with negativity.

\textit{Acknowledgment}: T. Pramanik thanks UGC, India for financial support. We thank Prof. A. Mazumdar and Prof. H. S. Sim for useful discussions.


\begin{thebibliography}{99}
\bibitem{rank} Schmidt decomposition \cite{ES} is a very good tool to describe composite systems. If $|\Psi\rangle$
is a pure state of composite systems A and B then,
$|\Psi\rangle =\sum_{i=1}^{d_A} \sqrt{\lambda_i} |i_A\rangle |i_B\rangle$,
represents the Schmidt decomposition of $|\Psi\rangle$, where ${|i_A\rangle} (i_A=1,2,...,d_A) \in\cal H_A$ and
${|i_B\rangle} (i_B=1,2,...,d_B) \in\cal H_B$ are orthonormal bases for A and B respectively, and $d_A\le d_B$. Here
$\lambda_i$ are the Schmidt coefficients,  non-negative real numbers  satisfying the relation
$\sum_{i} \lambda_i=1$. The number of non-zero Schmidt coefficients of a state is called its Schmidt rank. For bipartite mixed state ($\rho=\sum_{i} p_i|\Psi_i\rangle\langle\Psi_i| $) in finite dimension, the Schmidt rank is defined as \cite{Terhal} $SR(\rho)=\inf_{\{\sum_{i} p_i,|\Psi_i\rangle\}}\sup_i SR(|\Psi_i\rangle)$ where, the infimum is taken over all the ensembles $\{p_i,|\Psi_i\rangle \}$ of pure states with the average state $\rho$.
\end{thebibliography}
\end{document}